\documentclass[prb,twocolumn,superscriptaddress]{revtex4}

\usepackage{epsfig}
\def\figwidth{9cm}

\def\tp{t_\perp}
\def\kro{K_{\rho}}
\def\R{\mbox{Re}}

\def\SP{$(\mbox{TMTSF})_2\mbox{PF}_6$ }

\def\SX{$(\mbox{TMTSF})_2\mbox{X}$ }
\def\est{\epsilon^*}
\def\sig0{\sigma_{yy}^{(0)}(\omega)}
\def\rsig0{\sigma_{yy}^{(0)}(\omega)^\prime}
\def\isig0{\sigma_{yy}^{(0)}(\omega)^{\prime\prime}}
\def\cond0{\sigma_{yy}^{(0)}}

\begin{document}

\title{Hall effect and inter-chain magneto-optical properties of
coupled Luttinger liquids}

\author{Andrei Lopatin}
\email{lopatin@physics.rutgers.edu} \affiliation{Department of
Physics, Rutgers University, Piscataway, New Jersey  08854,
U.S.A.}

\author{Antoine Georges}
\email{antoine.georges@lpt.ens.fr} \affiliation{CNRS UMR 8549-
Laboratoire de Physique Th{\'e}orique, Ecole Normale
Sup{\'e}rieure, 24 Rue Lhomond 75231 Paris Cedex 05, France}
\affiliation{Laboratoire de Physique des Solides, CNRS-UMR 85002,
Universit\'e Paris--Sud, B\^at. 510, 91405 Orsay, France}

\author{T. Giamarchi}
\email{giam@lps.u-psud.fr} \affiliation{Laboratoire de Physique
des Solides, CNRS-UMR 85002, Universit\'e Paris--Sud, B\^at. 510,
91405 Orsay, France} \affiliation{CNRS UMR 8549- Laboratoire de
Physique Th{\'e}orique, Ecole Normale Sup{\'e}rieure, 24 Rue
Lhomond 75231 Paris Cedex 05, France}

\date{\today}

\begin{abstract}
We consider the  Hall effect in  a system of weakly coupled
Luttinger chains. We obtain the full conductivity tensor in the
absence of dissipation along the chains. We show that while the dependence
of the  Hall and transverse conductivities on temperature and frequency are
affected by the Luttinger interaction very strongly, the Hall
resistivity is given by a simple
formula corresponding to the noninteracting fermions. We compute the
frequency, temperature and field dependence of the transverse conductivity.
Consequences for the quasi-one-dimensional organic conductors are discussed.
\end{abstract}
\pacs{}

\maketitle

\section{Introduction} \label{intro}

Interacting electrons in one dimension form a non-Fermi liquid
state, usually called a Luttinger liquid (LL)
\cite{haldane_bosonisation}. There are several candidates for
actual realizations of this state in real systems, including
organic conductors, edge states and stripe phases in the Quantum
Hall effect, carbon nanotubes, etc... In many cases however, one
is dealing only with a {\it quasi} one-dimensional situation in
which the chains are weakly coupled to one another. In the
context of organic conductors in particular, the most relevant
type of inter-chain coupling is an inter-chain hopping $\tp$. The
question that immediately arises is  whether the Luttinger
liquid  effects can be observed in quasi-one dimensional systems.
The most naive answer would be that such systems are essentially
two-dimensional, and perhaps Fermi liquids, at energy scales
lower than the interchain hopping term $\tp$, but that at
energies larger than this hopping the Luttinger liquid effects
survive. The renormalization group calculation supports this
picture, but due to the effect of the interaction between
electrons, the inter-chain hopping term is strongly renormalized
\cite{bourbonnais_rmn} and the crossover between the
one-dimensional Luttinger liquid and two-dimensional behavior
takes place at the energy scale:
\begin{equation}\label{estar}
\epsilon^*=t_\perp (t_\perp/\epsilon_F)^{\eta/(1-\eta)}
\end{equation}
In this expression, $\epsilon_F$ is the Fermi energy of a single
chain ($\epsilon_F\propto t$, the in-chain hopping) and $\eta/2$
is the scaling dimension of the physical electron operator in the
Luttinger liquid. For a model with spin: $ \eta =
(\kro+1/\kro)/4-1/2$, with $\kro$ the LL parameter in the charge
sector. $\eta$ is also the exponent associated with the
singularity of $n(k)$ at the Fermi surface\footnote{$\eta$ was called
$\alpha$ in Ref.~{\protect\onlinecite{georges_organics_dinfiplusone}}
 and $\eta=2\Delta$
in Ref.~{\protect\onlinecite{lopatin_hall_luttinger}}}.
When interactions are strong ($\kro\ll 1$ i.e $\eta$ not too small),
$\est$ can be much smaller than the naive estimate $\est=\tp$,
suggesting that LL behavior could be observable down to a much
smaller energy (temperature) scale than the bare $\tp$.

We emphasize that the estimate (\ref{estar}) does not take into
account the effect of an intra-chain umklapp scattering. When
relevant, this coupling tends to open a Mott gap. One is thus
faced with the more complicated situation of two relevant
perturbations competing with each other, and this competition
controls the physics of the dimensional crossover. It was
recently pointed out\cite{georges_organics_dinfiplusone} that
this may be an important consideration for the physics of the
quasi one-dimensional organic conductors, even for the metallic
compounds \SX. A full understanding, in these
compounds\cite{jerome_organic_review}, of the
crossover between the high energy (temperature, frequency) phase,
likely to be indeed a Luttinger liquid
\cite{schwartz_electrodynamics,moser_conductivite_1d,%
henderson_transverse_optics_organics}, and the low energy one is
still lacking.

The present paper is devoted to the calculation of the inter-chain
conductivity and Hall effect in a system of weakly coupled LL
chains, in the presence of a magnetic field perpendicular to the
chains. We work to lowest order in a perturbative expansion in
the inter-chain hopping $\tp$, which is justified when one of the
characteristic energies associated with the temperature $T$, the
frequency $\omega$ or the field $H$ is larger than $\est$.
Furthermore, the band curvature ($\alpha$) must also be treated
as a perturbation since for a purely linear spectrum particle-hole
symmetry would lead to a vanishing Hall conductance. As explained
below, there are subtleties associated with these perturbative
expansions, having to do with the non-commutativity of the small
$\omega$ and small $\alpha$ or small $\tp$ expansions. Because of
these technical difficulties, we mainly focus in this paper on
the case where there is {\it no dissipation inside the chains} (in
particular, no umklapp scattering) -see, however, the remarks
made in the conclusion-. Two main results are obtained in this
paper: i) we show that in the absence of in-chain dissipation,
{\it the Hall resistance is a constant}, independent of
frequency and  temperature (this result is exact for the model that
we study) ii) using the perturbation theory in the interchain hopping
we derive an
explicit expression for the inter-chain conductivity as a
function of temperature, frequency and magnetic field
Eq.~(\ref{general_trans}), which also determines the full
resistivity tensor in this dissipationless limit
Eqs.~(\ref{rhoxx}-\ref{conduc_tensor}).

To make contact with previous works, we note that the inter-chain
conductivity in zero field was considered in
Refs.~\onlinecite{clarke_coherence_coupled,anderson_quasi1d_coherence,%
moser_conductivite_1d} and
recently reexamined in
Ref.~\onlinecite{georges_organics_dinfiplusone}.
The Hall conductivity was first considered in
Ref.~\onlinecite{lopatin_hall_luttinger} at zero
temperature and for very large magnetic fields. Here we
generalize this theory to finite temperature and arbitrary fields
(including the physically important small field limit).

There are several experimental motivations for the present work.
Inter-chain transport and optical measurements have been used as
a probe of the in-chain physics in quasi-one dimensional organic
conductors, and a critical discussion of the relevance of
coupled Luttinger liquid models to these measurements has been
given recently by two of the authors \cite{georges_organics_dinfiplusone}. Very recently,
measurements of the Hall effect in the Bechgaard salt \SP in two
different geometries have been reported \cite{moser_hall_1d,mihaly_hall_1d}.
While in-chain dissipation may be an important ingredient for the
understanding of these experiments, the results of the present
work for an ideal Luttinger liquid provide a benchmark to which the
experiments can be compared.

The paper is organized as follows: In Sec.~\ref{model}, we
introduce the model and define the geometry that we consider. In
Sec.\ref{gal}, we prove some general properties of the Hall
effect in the absence of dissipation in the chains. In
Sec.\ref{trcondac} we calculate the transverse conductivity, and
the optical Hall angle. The main results are summarized in
Sec.\ref{conclusion}, where we also briefly discuss the possible
consequences of in-chain dissipation. Technical details can be found
in the appendices.

\section{Model and Geometry}
\label{model}

\begin{figure}
\centerline{\epsfig{file=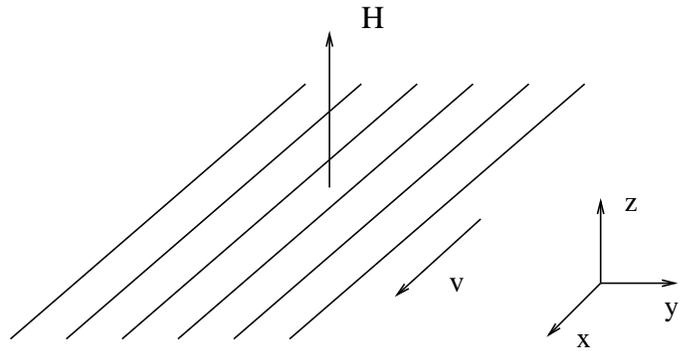,angle=0,width=\figwidth}}
\caption{\label{fig:geometry} This figure shows the orientation of
the magnetic field and coordinate frame with respect to the
system of coupled chains. Considering this system from the
reference frame that moves along the x-direction allows to obtain
dc- Hall resistivity and conductivity.}
\end{figure}

The geometry considered in this paper is depicted in
Figure~\ref{fig:geometry}.
We consider one-dimensional chains (along the x-axis), coupled by
a transverse hopping (y-axis) into a two-dimensional array. A
magnetic field is applied perpendicular to the array (z-axis).
For simplicity we  consider the case of spinless electrons, and
neglect the in-chain umklapp processes (a legitimate assumption
if the system is not at a commensurate filling). The main
technical difficulty of this problem is  that the Hall
conductivity of chains with  a linear  electron spectrum
\begin{equation}
\epsilon_\pm=\pm v_F(p \mp p_F)
\end{equation}
is zero due to the particle-hole symmetry. Therefore to obtain a
nontrivial answer it is necessary to consider a nonlinear
correction to the spectrum
\begin{equation}\label{alphasp}
\epsilon_\pm=\pm v_F(p\mp p_F)+\alpha (p\mp p_F)^2.
\end{equation}
Thus the Hamiltonian of the problem is
\begin{eqnarray}
H=\int  dx\Biggl[\sum_{i}v_F\hat\psi_i^\dagger\tau_3
(-i\partial_x)\hat\psi_i-
\alpha \sum_{i} \hat\psi^\dagger_i\partial_x^2\hat\psi_i \nonumber \\
+g\sum_i\hat\psi_{i+}^\dagger\hat\psi_{i+}\hat\psi_{i-}^\dagger\hat\psi_{i-}
-t_\perp \sum_{\langle i,j \rangle} \hat\psi^\dagger_i\hat\psi_j
e^{-i{e\over c}A_{i,j}}\Biggr] \label{ham0},
\end{eqnarray}
where $\hat\psi$ is a two-component vector composed from the
right- and left-moving electrons $\hat\psi=\left(\begin{array}{c}
\hat\psi_+\\ \hat \psi_- \end{array}\right)$, $\tau_3$ is a Pauli
matrix and $A_{i,j}=\int_i^{j}{\bf A} d{\bf l}$.  We  use the
Landau gauge $A_y=Hx$. The second term in the Hamiltonian
(\ref{ham0}) corresponds to the nonlinear correction
(\ref{alphasp}) to the free electron spectrum. The model without
the hopping term and $\alpha-$term can be solved exactly, for
example by the bosonization method. It is also possible  to
bosonize the  $\alpha$-term \cite{haldane_bosonisation} but it
leads to a model with a cubic interaction between the bosons
which is not exactly solvable. The perturbation theory in 
$\alpha-$term seems to be always a good approximation
because the nonlinear term in the spectrum
(\ref{alphasp}) is small compared with the linear one as long as
the typical energy scale of the problem is less than the Fermi
energy. This is true, indeed, but only for the quantities that 
have  a regular expansion in $\alpha.$  As we show below, the components 
of the resistivity tensor depend regular on $\alpha$ and therefore
can be expanded in $\alpha.$ On the contrary, the components of the 
conductivity tensor at low frequencies depend singular on $\alpha$ 
and, thus, cannot be expanded in $\alpha.$

We consider in this paper the calculation of the anisotropic
conductivity tensor (in the presence of the magnetic field),
relating the current to the electric field:
\begin{equation}
\left[\begin{array}{cc} j_x \\j_y \end{array}\right]=
\left[\begin{array}{cc} \sigma_{xx} & \sigma_{xy} \\
  \sigma_{yx} & \sigma_{yy}\end{array}\right]
\left[\begin{array}{cc}E_x \\ E_y\end{array} \right].
\end{equation}
The resistivity tensor is equal to the inverse of the conductivity
tensor, and the Hall resistivity $\rho_{xy}$ is in particular given
by:
\begin{equation}
\rho_{xy}={{-\sigma_{xy}}\over{\sigma_{xx}\sigma_{yy}+\sigma_{xy}^2}}.
\end{equation}

In the following, we shall evaluate the conductivity and
resistivity tensor using a perturbative method in both the
band curvature $\alpha$ and the inter-chain hopping $\tp$.
It is crucial to realize that the limits of small
$\alpha$, small $\tp$ and small frequency $\omega$
do not commute. As a result, great care must be taken in
order to decide which quantity to expand. This will be fully
clarified in the next section.

\section{Hall effect in the absence of in-chain dissipation:
general considerations} \label{gal}

In this section we show that, in the case where
there is no dissipation in the chains,
several aspects of the Hall effect
can be deduced from general principles.
We first consider (Sec.\ref{sec:galilee})
the limit where $\omega$ is taken to zero
first, in which Galilean invariance can be used.
We then turn  to non-zero frequency (Sec.~\ref{commutation})
and use the commutation relation of the in-chain current operator
and Hamiltonian  in order to
obtain general expressions for the conductivity and
resistivity tensors. These expressions involve a single
non-trivial quantity, the inter-chain resistance
$\rho_{yy}(\omega,T)$ which will be explicitly evaluated
(to lowest order) in Sec.\ref{trcondac}. They also allow us to
clarify which quantities have regular perturbative expansions in
$\alpha$ and $\tp$. In Sec.\ref{sec:pulse}, we explain physically
the content of these expressions.

\subsection{Galilean invariance and
the $\omega\rightarrow 0$ limit} \label{sec:galilee}

Here we focus on the zero-frequency limit, and show
that when the Hamiltonian is Galilean invariant in the  direction
of the chains, the Hall resistance is independent of temperature
and simply given by its free electron value. The resistivity
tensor is defined by
\begin{equation}
\left[\begin{array}{cc}E_x \\ E_y \end{array} \right]=
\left[\begin{array}{cc} \rho_{xx} & \rho_{xy} \\
 \rho_{yx} & \rho_{yy} \end{array}\right] \left[\begin{array}
{cc}j_x\\ j_y \end{array} \right]. \label{resmat}
\end{equation}
To find the Hall resistivity $\rho_{xy}$ it is convenient to
apply the electric field  in such a way that the current will
flow  exactly along the chains, then $j_y=0$ and
(\ref{resmat}) gives
\begin{eqnarray}
E_y &=& \rho_{yx}\; j_x, \label{rhoxy} \\
E_{x} &=& \rho_{xx}\;j_x.
\end{eqnarray}
Let us furthermore impose that $E_x=0$ (e.g. by imposing periodic conditions
along $x$).

We now envision a different
setup in which no external electric field and no current are
applied, and consider a reference frame that moves with respect
to this system at a velocity $v$ along the $x$ axis. In this
moving frame, $j_y$ is still vanishing because the transverse
current is  not affected by the Galilean transformation.
Due to such a Galilean transformation $E_x$ is still zero but
an electric field is induced along the y-axis, given by:
\begin{equation} \label{eq:ey}
E_y = - v {H\over c}
\end{equation}
Finally, in the moving frame, there is a  current along the
chains, which can be evaluated as follows. The velocity of right
(left) moving modes close to the Fermi surface is given by: $v_k
= \pm v_F + 2\alpha (k\mp k_F)$, so that a constant velocity $v$
corresponds  to a momentum shift $\Delta k = \frac{v}{2\alpha}$.
This in turn corresponds to a (one -dimensional) current density
in a single chain given by: $j_{1D}= - e v_F \Delta k/\pi$.
Finally, we obtain the induced current density along $x$ as:
\begin{equation} \label{eq:jx}
j_x = \frac{j_{1D}}{a_y} = - \frac{e v_F}{2\pi\alpha a_y}\, v
\end{equation}
Combining the above equations for $E_y$ and $j_x$, we finally
obtain:
\begin{eqnarray}
\rho_{xx}(\omega=0) &=& 0 \\
\rho_{yx}(\omega=0) &=& \frac{E_y}{j_x} = 2\pi\alpha\, {{H a_y}\over{v_F e
c}} \label{dchr}
\end{eqnarray}
This expression for the Hall resistivity can also be written in terms
of the electron density $n$, which is related to the in-chain Fermi momentum
through the relation: $na_y = k_F/\pi$ (since $na_y$ is the
density per one chain). Therefore:
\begin{equation}
\label{rhoyx}
\rho_{yx} = \frac{H}{nec} \, \frac{2\alpha k_F}{v_F}
\end{equation}
In the case of a purely parabolic band $\epsilon_k =
\frac{k^2}{2m}$, one has $2\alpha =1/m$ and $v_F=k_F/m$, so that
one recovers the familiar expression:
\begin{equation}
\rho_{yx} = \frac{H}{nec}
\end{equation}
For a tight-binding dispersion along the chains $\epsilon_k = -2t
\cos k$, one has $v_F=2t\sin k_F$ and $\alpha = t \cos k_F$, so
that:
\begin{equation}
\label{cosine}
\rho_{yx} = \frac{H}{nec} \, \frac{k_F}{\tan k_F}
\end{equation}
Hence, we have seen that Galilean invariance implies that the
dc- Hall resistivity is T-independent and unchanged by interactions.

In a similar way one can use the conductivity tensor
\begin{equation}
\left[\begin{array}{cc}j_x \\ 0  \end{array}\right]=
\left[\begin{array}{cc} \sigma_{xx} & \sigma_{xy} \\ -\sigma_{xy}&
\sigma_{yy}
\end{array}\right]
\left[\begin{array}{cc} 0 \\ E_y \end{array}\right],
\end{equation}
where $j_x$ and $E_y$ are again given by (\ref{eq:jx}) and (\ref{eq:ey}).
This leads to
\begin{eqnarray}
\sigma_{xy}(\omega=0)&=& \rho_{yx}(\omega=0)^{-1}, \\
\sigma_{yy}(\omega=0)&=& 0.
\end{eqnarray}
Thus the Galilean invariance implies, at zero frequency
\begin{eqnarray} \label{eq:galilee}
\rho_{xx}(\omega=0)=0,\quad
\sigma_{yy}(\omega=0)=0 \\
\sigma_{xy}(\omega=0)^{-1}=
-\rho_{xy}(\omega=0)
=2\pi\alpha {{H a_y}\over {v_F e c}}.
\end{eqnarray}
Note that $\rho_{xx}=0$ is well in agreement with our hypothesis of a perfect
conductor in the chain direction. (\ref{eq:galilee}) also shows that at low
frequency, the Hall {\it resistivity} is perturbative in the band curvature,
while $\sigma_{xy}$ is not. Finally, we emphasize that the fact
that $\sigma_{yy}(\omega=0)= 0$ does not provide any information
on the interchain {\it resistivity} $\rho_{yy}(\omega=0)$ which is
a finite quantity (at finite temperature) that must be calculated
independently (Sec. \ref{trcondac}).

\subsection{Commutation relations and general 
expressions for the resistivity tensor}
\label{commutation}

In order to go beyond the zero frequency limit let us compute
the commutator of the current along the chains
\begin{eqnarray} \label{eq:comcur}
\hat J_x&=&\int dx \sum_i \hat j_x(x,i) \\
\hat j_{x}(x,i)&=&{1\over {a_y}}
\left(ev_F\;\hat\psi^\dagger_i\tau_3 \hat\psi_i +2 e \alpha\,
\hat\psi^\dagger_i(-i\partial_x)\hat\psi_i\right),
\end{eqnarray}
with the Hamiltonian (\ref{ham0}). This gives  the remarkable result
\begin{equation}
\dot{\hat J_x}\equiv i[\hat{H}, \hat J_x]=\gamma \hat J_y,
\label{comrel}
\end{equation}
where
\begin{eqnarray} \label{eq:intercurrent}
\hat J_y&=&\int dx \sum_i \hat j_{y}(x,i), \\
\hat j_{y}(x,i)&=&t_\perp ei\left(\hat\psi^\dagger_i(x)\hat\psi_{i+1}(x)
e^{-i{e\over c}A_{i,i+1}}-h.c.\right),
\end{eqnarray}
is the current perpendicular to the chains and
\begin{equation}
\gamma = {{2e\alpha H }\over{c}}.
\end{equation}

 The commutation relation (\ref{comrel})
allows to find general relations between the components of the 
conductivity tensor: Suppose that a short pulse of the electric field 
$E_y=E_0 \delta(t)$
is applied along the y-axis, then
\begin{eqnarray}
j_x(t)=E_0\, \sigma_{xy}(t) \\
j_y(t)=E_0\, \sigma_{yy}(t)
\end{eqnarray}
and using (\ref{comrel}) we get
\begin{equation}
\dot \sigma_{xy}(t)=
\gamma\, \sigma_{yy}(t).
\end{equation}
Using this equation we obtain the relation between 
$\sigma_{yy}(\omega)$ and $\sigma_{xy}(\omega):$
\begin{eqnarray}
&&\gamma\,\sigma_{yy}(\omega)
=\gamma\int_0^\infty \sigma_{yy}(t) e^{i\omega t} dt= \nonumber \\
&&\int_0^\infty e^{i\omega t} \dot \sigma_{xy}(t)=-\sigma_{xy}(t=0^+)-i\omega
\sigma_{xy}(\omega). \label{sxysyy}
\end{eqnarray}
 From the Galilean invariance principle we know that $\sigma_{yy}(\omega=0)=0$
and $\sigma_{xy}(\omega=0)$ is finite, therefore from Eq.(\ref{sxysyy})
it follows that $\sigma_{xy}(t=0^+)=0$ and finally we obtain 
\begin{equation}
\sigma_{xy}(\omega)={{\gamma}\over{-i\omega}} \sigma_{yy}(\omega).
\label{rel1}
\end{equation}
Moreover from the Galilean invariance we have that
 $\sigma_{xy}(0)=v_F e c/2\pi\alpha H a_y,$ therefore
\begin{equation}
\sigma_{yy}(\omega)=
-{{v_F e^2}\over{\pi a_y}} {{i\omega}\over{\gamma^2}}, \;\;\;\;
\omega\to 0.     \label{sigmaxy0}
\end{equation}

 A similar consideration using the commutation relation (\ref{comrel}) 
when a short electric pulse is applied along the x-axis leads to
\begin{equation}
\sigma_{xx}(\omega)=
{{\gamma\,\sigma_{xy}(\omega)-\sigma_{xx}^D}\over{i\omega}}
\label{sigmaxx}
\end{equation}
where
\begin{equation}
\sigma_{xx}^D={{e^2 v_F }\over{\pi a_y}}.
\end{equation}
which is consistent with the expected large frequency behavior of
$\sigma_{xx}$ (perfect conductor).
Now, using (\ref{rel1}) $\sigma_{xx}$ may be related with $\sigma_{yy}$ as
\begin{equation}
\sigma_{xx}(\omega)=
-{{e^2 v_F }\over{\pi a_y }}{1\over{i\omega}}+{{\gamma^2}\over
{\omega^2}} \sigma_{yy}(\omega). \label{sigmax}
\end{equation}
According to (\ref{sigmaxy0}) we see
that the singularities $1/\omega$ in (\ref{sigmax}) are canceled
out and that, in contrast to naive intuition, the {\it conductivity} $\sigma_{xx}$
is finite at zero frequency !
This is of course due to the presence of the magnetic field, and occur
for frequencies smaller than the ``cyclotron frequency'' of the anisotropic
system. The above formulas show that at low frequency the conductivity
tensor is drastically affected by the presence of the magnetic field an thus
does not allow for perturbative calculations.
The resistivity tensor, in contrast, has a
well-behaved expansion in the band curvature.

In order to derive expressions for this tensor, we use
(\ref{rel1},\ref{sigmaxx}) to obtain:
\begin{equation}
\sigma_{xx}(\omega) \sigma_{yy}(\omega)
+\sigma_{xy}^2(\omega)={1\over \gamma}\, \sigma_{xx}^D\,\,
\sigma_{xy}(\omega)
\end{equation}
Therefore $\rho_{xx}$ is given by
\begin{equation}
\rho_{xx}(\omega)={{
\sigma_{yy}(\omega)}\over{\sigma_{xx}(\omega) \sigma_{yy}(\omega)
+\sigma_{xy}^2(\omega)
}}=-{{i\omega}\over{\sigma_{xx}^D}}=-{{\pi a_y }\over{v_F e^2 }}\,\,i\omega.
\label{rhoxx}
\end{equation}
We see that $\rho_{xx}$ is independent of magnetic field, inter-chain hopping and band
curvature, and that $\rho_{xx}(0)=0$ in  agreement with the result following
from Galilean invariance.
We obtain $\rho_{xy}(\omega)$ in a similar manner:
\begin{eqnarray}
\rho_{xy}(\omega) &=&
{{-\sigma_{xy}(\omega) }\over{\sigma_{xx}(\omega)\sigma_{yy}(\omega)
+\sigma_{xy}^2(\omega) }} \nonumber \\
&=& -2\pi\alpha
{{H a_y }\over{v_F ec}}\equiv -\gamma {{a_y \pi}\over{v_F e^2}},
\end{eqnarray}
This generalizes to finite frequency the result previously obtained
using Galilean invariance. $\rho_{xy}(\omega)$ is seen to be independent of
frequency and temperature. This result established here from general
principles can also be recovered from an explicit calculation
valid at high frequency,
presented in appendix~\ref{hallcon}.

We note that these general considerations do not allow for the
determination of the component $\rho_{yy}(\omega,T;H)$ of the
resistivity tensor. This is the only non-trivial quantity that
must be calculated explicitly in this dissipationless case. It is
proven in Appendix ~\ref{ap:expan} that this quantity has a well-behaved expansion
in powers of the band curvature parameter $\alpha$, so that to
lowest order one can use the value of $\rho_{yy}$ for $\alpha=0$.
To this zeroth order, the inter-chain conductivity and
resistivity are simply related by:
\begin{equation}
\rho_{yy}(\omega,T;H)^{(0)} = 1/ \sigma_{yy}(\omega,T;H)^{(0)}
\end{equation}
In Sec.\ref{trcondac}, an expansion of $\sigma_{yy}(\omega,T;H)^{(0)}$ to lowest
order in the inter-chain hopping is performed (i.e to order
$\tp^2$), which thus fully determines $\rho_{yy}(\omega,T;H)^{(0)}$
to lowest order.

Finally, for the sake of completeness, we give the general
expression of the conductivity tensor in terms of
$\rho_{yy}(\omega)$:
\begin{eqnarray}
\sigma_{xx}(\omega)={{e^2 v_F}\over{\pi a_y}}\,{{1 }\over{-i\omega
+\gamma^2{{a_y\pi}\over{v_F e^2}}\rho_{yy}^{-1}(\omega)}} \nonumber\\
\sigma_{xy}(\omega)={\gamma \over{-i\omega\rho_{yy}(\omega)+
{{a_y\pi}\over{v_F e^2}}\gamma^2 }} \label{conduc_tensor}\\
\sigma_{yy}(\omega)= {1\over{ \rho_{yy}(\omega)-
{{\gamma^2}\over{i\omega}} {{a_y \pi}\over{v_F e^2}}}} \nonumber
\end{eqnarray}

From the equation relating $\sigma_{yy}(\omega)$ and $\rho_{yy}(\omega)$ we
see that at high  enough frequency the effect of curvature 
becomes not important and 
$\rho_{yy}^{-1}(\omega)=\sigma_{yy}(\omega)=\sigma_{yy}^{(0)}(\omega).$
Assuming that $\sigma_{yy}^{(0)}$ is finite at zero frequency
(this is the case for any finite temperature) we  estimate
the crossover frequency below which the curvature effects become
important
\begin{equation}
\omega_0 ={{\gamma^2\,\pi
  a_y \sigma_{yy}^{(0)}(\omega=0)}\over{v_F\, e^2}}. \label{omegazero}
\end{equation}
Thus for frequencies higher than $\omega_0$ one can neglect the curvature
effects and use the zero curvature result for transverse conductivity 
calculated in Sec.\ref{trcondac} directly.
 At frequencies lower than $\omega_0$ the zero curvature result may be used only 
for resistivity $\rho_{yy},$  since this quantity depends regularly on 
$\alpha$ even at low frequencies (see Appendix\ref{ap:expan}).   
Thus, at frequencies lower than $\omega_0$  the conductivity tensor 
should be obtained by inserting in the expressions (\ref{conduc_tensor}) the
result for $\rho_{yy}(\omega,T;H)^{(0)} = 1/\sigma_{yy}(\omega,T;H)^{(0)}$ 
calculated in Sec.\ref{trcondac}.

\subsection{Physical arguments:
response to a current pulse and Hall angles}
\label{sec:pulse}

Here, we rederive some of the previous results using physical arguments, and
give expressions for the Hall angles of this anisotropic system.
Following Ref.~\onlinecite{drew_hall_twoscattering} (see also
Ref.~\onlinecite{lange_hall}), we consider the following thought
experiment. A current pulse $j_y(t) = j^y_0 \delta(t)$ is sent
into the system along the y-direction (i.e. perpendicular to the
chains and to the magnetic field). A transient current $j_x(t)$
is thus induced along the chains, and we imagine that the
electric field along the chain is maintained to $E_x=0$. Using
the definition of the conductivity tensor, the transient current
$j_x(t)$ is found to be (for $t>0$):
\begin{equation}
j_x(t) = j^y_0 \tan\theta_H^x(t)
\end{equation}
in which $\theta_H^x(t)$ is the Fourier transform
\begin{equation}
\tan\theta_H^x(t) = \int \frac{d\omega}{2\pi} e^{-i\omega t}
\tan\theta_H^x(\omega+i0^+)
\end{equation}
of the (retarded) frequency-dependent Hall angle:
\begin{equation}
\tan\theta_H^x(\omega) =
{{\sigma_{xy}(\omega)}\over{\sigma_{yy}(\omega)}}.
\end{equation}
In the absence of dissipation along the chains, we expect no time
decay of the induced current $j_x(t)$, i.e $j_x(t)=j_x(t=0)=C_x j^y_0
\theta(t)$ with $C_x$ a constant. ($C_x$ could a priori be temperature dependent,
but the above reasoning using Galilean invariance shows that it is not). This
implies that:
\begin{equation}
\tan\theta_H^x(\omega) = C_x \frac{i}{\omega}
\end{equation}
and hence that $\sigma_{xy}$ and $\sigma_{yy}$ are simply
proportional at all frequencies and temperatures:
\begin{equation}
\sigma_{xy}(\omega,T) = C_x \frac{i}{\omega+i0^+}
\sigma_{yy}(\omega,T) \label{sigprop}
\end{equation}
one thus recovers the relation (\ref{rel1}), which also determine
the constant $C_x$ to be $C_x=2\alpha e H/c$

One can also consider the complementary thought experiment in
which a current pulse $j_x(t)=j_0^x \delta(t)$ is generated along
the chains. The relaxation of the induced Hall current
perpendicular to the chains (with the constraint $E_y=0$) is then
given by:
\begin{equation}
j_y(t) = j^x_0 \tan\theta_H^y(t)
\end{equation}
with similar notations as above, and the optical Hall angle
$\theta_H^y$ given by:
\begin{equation}
\tan\theta_H^y(\omega) =
{{\sigma_{yx}(\omega)}\over{\sigma_{xx}(\omega)}} =
\frac{\rho_{xy}}{\rho_{yy}(\omega)}
\end{equation}
To lowest order (in the band curvature and inter-chain hopping),
we can thus
relate this Hall angle to the inter-chain conductivity as follows:
\begin{equation} \label{eq:trivial}
\tan\theta_H^y(\omega) =
\rho_{xy}\sigma_{yy}^{(0)}(\omega) + \cdots
\end{equation}
Using the spectral representation $\sigma(z)=\frac{i}\pi
\int d\epsilon \frac{\R \sigma(\epsilon)}{z-\epsilon}$,
one obtains the decay of the
current pulse along $y$ as:
\begin{equation}
j_y(t) = \rho_{xy} j_0^x \theta(t)
\,\int_{-\infty}^{+\infty} \frac{d\omega}{2\pi} e^{-i\omega t}
\R \,\sigma_{yy}^{(0)}(\omega)
\end{equation}
In the next section, an explicit expression will be obtained for the
frequency, temperature, and magnetic field dependence of
$\sigma_{yy}^{(0)}$, hence allowing the determination of the time
decay of the current $j_y(t)$.

\section{Transverse conductivity}
\label{trcondac}

In this section, we calculate the transverse
conductivity $\sigma_{yy}^{(0)}$ in the absence of band curvature, to lowest
order in the inter-chain hopping, as a function of frequency, temperature and magnetic
field. As discussed above, this then completely determines the
conductivity tensor (and Hall angle) by inserting
$\rho_{yy}^{(0)}=1/\sigma_{yy}^{(0)}$ in the expressions
(\ref{conduc_tensor}).

According to the Kubo formula, the conductivity is given by
\begin{equation}
\sigma_{yy} (\omega)= \sigma_{yy}^{\cal
P}(\omega)+\sigma_{yy}^{\cal D}(\omega),
\end{equation}
where $\sigma^{\cal P}(\omega)$ and $\sigma^{\cal D}(\omega)$
are the paramagnetic  and
the diamagnetic  contributions respectively. The paramagnetic term is
\begin{equation}
\sigma_{yy}^{\cal P}(\omega)={{a_y}\over\omega} \sum_i\int dx\,
P_R(x,i,\omega)\label{sigmaom},
\end{equation}
where $P_R$ is the retarded current-current correlator
\begin{equation}
P_R(x,i,\omega)= \int_{0}^{\infty}dt\,\langle e^{i\omega t} [\hat
j_y (x,i,t),\hat j_y(0,0,0)]\rangle,
\end{equation}
and $\hat j_y$ is the operator of the inter-chain current (\ref{eq:intercurrent}).

The diamagnetic term is given by
\begin{equation}
\sigma_{yy}^{\cal D}={{-e^2\,t_\perp a_y}\over{i\,\omega}}\,
\left\langle \hat\psi^\dagger_0(0)\hat\psi_1(x)\,e^{-ie
A_{0,1}/c}+c.c. \right\rangle.
\end{equation}
The retarded current-current correlator $P_R(\omega)$ can be
obtained from the Matsubara correlator
\begin{equation}
P_{\cal M}(x,i,\Omega)= \int_{0}^{\beta}d \tau \, e^{i\Omega
\tau}\langle T_\tau \, \hat j_y (x,i,\tau)\;\hat
j_y(0,0,0)\rangle  \label{pm}
\end{equation}
by the analytical continuation of the Matsubara frequency $\Omega$
to the real frequency $\omega$
\begin{equation}
P_R(\omega)=-iP_M(\Omega)_{i\Omega\to \omega+i0^+}.
\end{equation}
To the lowest order in $t_{\perp}$ the paramagnetic term is
simply given by the expectation value of (\ref{pm}) with respect
to the single chain Hamiltonian. For the diamagnetic term one
should consider the hopping term in the Hamiltonian (\ref{ham0})
to the first order. Combining the lowest order expressions for
paramagnetic and diamagnetic terms we get
\begin{eqnarray}
\sig0 = {{2 t_\perp^2\,a_y e^2 }\over{i \omega}}
\sum_{s=\pm}\Biggl[
\int dx \int_0^{\beta} d\tau\; e^{i\Omega \tau}\; G_s^2(x,\tau) \nonumber \\
\times \cos(h x) -(\Omega=0)\Biggr]_{i\Omega\to\omega+i0^+}
\label{trcond}
\end{eqnarray}
In this expression
\begin{equation}
h=e H a_y/c
\end{equation}
is a characteristic energy scale associated with the magnetic
field, which will play an important role in the following.
$G_{\pm}$ are the single chain Green's function for each chiral
mode:
\begin{eqnarray}
G_{\pm}&&(x,\tau)={i\over{2}}\;\; {{T}\over {\sinh\pi T(\pm
x+i\tau) }}
  \nonumber \\
&\times&{{(\pi T  a_x)^{\eta}}\over[{\sinh(\pi T(x-i\tau))
\sinh(\pi T(x+i\tau))] ^{\eta/2} }}. \label{tgreenf}
\end{eqnarray}
The integral over $\tau$ in (\ref{trcond}) must be calculated for
the Matsubara frequency $\Omega,$ and then the analytical
continuation should  be taken. This is performed explicitly in
Appendix~\ref{appx_trans} in which the following general
expression is derived:
\begin{eqnarray}
\label{general_trans}
\rsig0 &=&{{ a_y e^2
t_\perp^2(2\pi T a_x)^{2\eta}}\over {2\pi^2}} {{\sinh
(\omega/2T)}\over {\omega}}  \nonumber
{{(\frac\eta2)^2+{{\omega^2+h^2}\over{(4\pi T)^2}} }\over
{\Gamma(\eta)\Gamma(2+\eta)}}
\nonumber \\
&&\left|\Gamma\left(\frac{\eta}{2}+{{\omega+h}\over{4\pi T}}i\right)
\right|^2 \left|\Gamma\left(\frac{\eta}{2}+{{\omega-h}\over{4\pi
T}}i\right) \right|^2.
\end{eqnarray}
In the following, we examine various limiting cases of physical
interest of this expression.

\subsection{Zero temperature} \label{zerotem}

At zero temperature the integrals in  (\ref{trcond}) can be
taken  analytically giving
\begin{equation}
 \sig0 = A(\eta)\,{1\over {i\omega}}\, \Bigl[(h^2-\omega^2)^{\eta}\,
{{h^2+\omega^2}\over{h^2-\omega^2}} -h^{2\eta} \Bigg],
\label{zerotemcon}
\end{equation}
where the coefficient $A(\eta)$ is
\begin{equation}
A(\eta)={{t_\perp^2 a_y\, e^2 }\over \pi} \left( { a_x\over 2}
\right)^{2\eta}{{\Gamma(1-\eta)}\over{\Gamma(2+\eta)}}.
\end{equation}
In the limit of low frequencies $\omega \ll h$ we get
\begin{equation}\label{zomega}
\sig0 = -A(\eta)(2-\eta)i\,\omega\,
\left(\frac{e H a_y}{c}\right)^{2\eta-2}.
\end{equation}
and thus a purely reactive response. For zero magnetic field
($h=0$) we have
\begin{eqnarray}
\rsig0 &=& A(\eta)\,\sin(\pi\eta)\,\,|\omega|^{2\eta-1}, \\
\isig0 &=&A(\eta)\,\cos(\pi\eta)\,\,|\omega|^{2\eta}/\omega,
\end{eqnarray}
where $\rsig0$ and $\isig0$
are the real and imaginary parts of the conductivity respectively.
 We see that the real and
imaginary parts of conductivity have the same dependence on
$\omega,$ so the ratio
$\rsig0/\isig0$ is $\omega$-independent:
\begin{equation}
{\rm sign} (\omega)\;\frac{\rsig0}{\isig0} = \tan(\pi\eta)
\end{equation}
This equation suggest an independent way to measure the  exponent
$\eta$ experimentally. Note that in case of Fermi liquid the
limit of the above ratio at $\omega\to 0$ should be zero.
Thus the conductivity dependence on $\omega$ reflects the
non-Fermi liquid properties of the in-chain Luttinger liquid.

When the magnetic field is applied, according to
(\ref{zerotemcon}), at frequencies  $|\omega|<h$  the real part
of the conductivity is zero and dissipation is absent. But when
$|\omega|>h$ the real part is not zero and it is given by
\begin{equation}
\rsig0 = A(\eta)\,{{\sin(\pi\eta)}\over{|\omega|}}
\,{{\omega^2+h^2}\over{(\omega^2- h^2)^{1-\eta}}}.
\end{equation}
Thus the magnetic field turns off dissipation for frequencies in
the region $|\omega|<h.$

\subsection{DC transverse conductivity}

\begin{figure}
\centerline{\epsfig{file=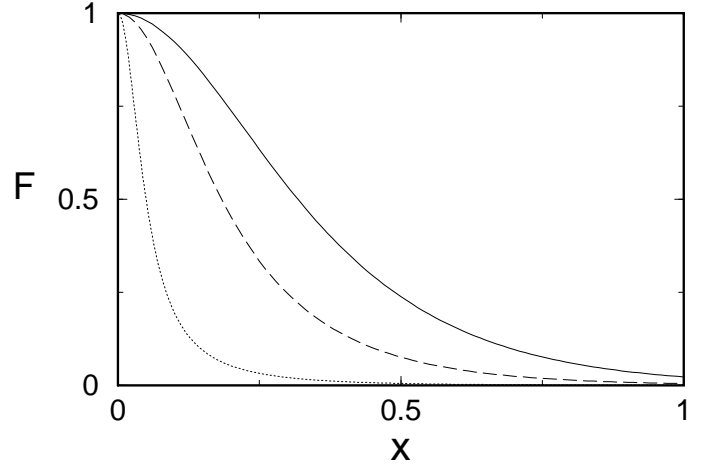,angle=0,width=\figwidth}}
\caption{\label{fig:hallmag} The function $F(\eta,x)$ that
determines dependence of the conductivity on the magnetic field
$(x=h/4\pi T)$ is plotted for different values of $\eta$: the
solid line corresponds to $\eta=0.8$, the dashed line to
$\eta=0.4$ and the  dotted line to $\eta=0.1$.}
\end{figure}

As shown in Appendix~\ref{appx_trans}, the d.c. transverse
conductivity can also be computed analytically. The imaginary
part is zero, and the real part is given by
\begin{eqnarray}
\cond0(T;H) &=&{{ a_y e^2 t_\perp^2(2\pi T a_x)^{2\eta}}\over
{4\pi^2 T}} {{(\frac\eta2)^2+h^2/(4\pi T)^2}\over
{\Gamma(\eta)\,\Gamma(\eta+2) }} \nonumber \\
& &|\Gamma\left(\eta /2 +h i/4\pi T\right)|^4. \label{dccon}
\end{eqnarray}
In the absence of the magnetic field we have
\begin{equation}
\cond0(T;H=0) ={{ a_y e^2 t_\perp^2(2\pi a_x)^{2\eta}}\over
{4\pi^2}}{{(\frac\eta2)^2\,\Gamma^4(\frac\eta2)}\over
{\Gamma(\eta)\,\Gamma(\eta+2)}}\; T^{-1+2\eta},
\end{equation}
so the conductivity depends on the temperature as
$T^{-1+2\eta}$, in agreement with the previous results of
two of the authors in
Ref.~\onlinecite{georges_organics_dinfiplusone}. If $\eta\ll
1$ one can simplify the above expression
\begin{equation}
\cond0 = {{ a_y e^2 t_\perp^2}\over {\pi^2}}{1\over{T\eta}}.
\end{equation}
Thus the system becomes a perfect conductor when $\eta\to 0$
as it should since the current commutes with the Hamiltonian. To
plot the dependence of the DC conductivity (\ref{dccon}) on the
magnetic field it is convenient to rewrite (\ref{dccon}) in the
form
\begin{eqnarray}
\cond0 = {{ a_y e^2 t_\perp^2(2 \pi  a_x)^{2\eta}}\over
{16 \pi^2 }} &&{{\eta^2\,\Gamma^4(\frac\eta2)}\over
{\Gamma(\eta)\,\Gamma(\eta+2)}} \nonumber \\
&&\times\,\, T^{-1+2\eta}\,\,\,F(\eta, h/4\pi T),
\end{eqnarray}
where the function $F$ is
\begin{equation}
F(\eta,x)= {{|\Gamma(\eta /2+i
x)|^4}\over{\Gamma^4(\frac\eta2)}}\;
{{\eta^2+4 x^2}\over{\eta^2}} \label{ffun}.
\end{equation}
The function $F(\eta,x)$ is plotted on
Figure~\ref{fig:hallmag} for different values of $\eta$.

In case  of large magnetic fields $h\gg T$ one can simplify
(\ref{ffun}) obtaining for conductivity the following expression
\begin{equation}
\cond0 =
{{a_y\,e^2\,t_\perp^2\,(
a_x/2)^{2\eta}}\over{\Gamma(\eta)\,
\Gamma(\eta+2)}}\,\,{{h^{2\eta}}\over T} \,e^{-h/2T}.
\label{zT}
\end{equation}
This is different from the result (\ref{zomega}) obtained for the
zero temperature case.  Thus the answer for the conductivity
depends on which limit is taken first: $\omega\to 0$ or $T\to 0.$
The answer (\ref{zomega}) corresponds to the limit $\omega \gg
T,\;\;\omega,T\to 0$ and the answer (\ref{zT}) to $T \gg
\omega,\;\;\omega,T\to 0.$

\subsection{Real part of the transverse AC conductivity}

\begin{figure}
\centerline{\epsfig{file=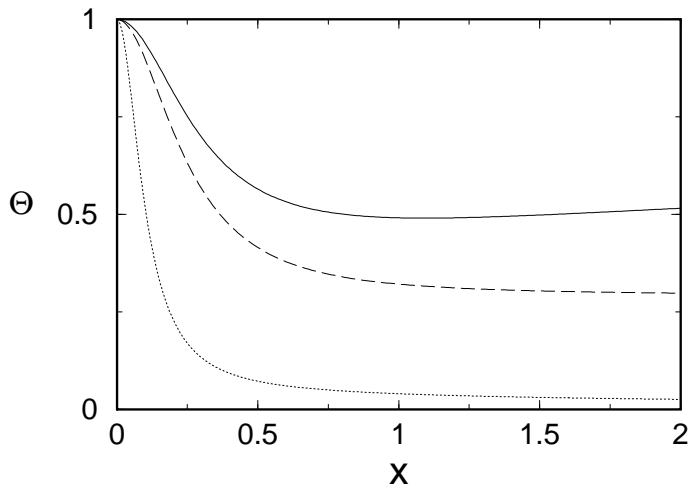,angle=0,width=\figwidth}}
\caption{\label{fig:hallcon} The function $\Theta(\eta,x)$
showing the dependence of the real part of the conductivity on
the frequency $\omega$ ($x=\omega/4\pi T)$ is plotted for
different values of $\eta$ the solid line corresponds to
$\eta=0.6$, the dashed line to $\eta=0.5$ and the  dotted
line to $\eta=0.2$.}
\end{figure}

In the absence of magnetic field the real part of the
conductivity can be written as (see Appendix~\ref{appx_trans})
\begin{eqnarray}
\rsig0 = {{ a_y e^2 t_\perp^2(2 \pi  a_x)^{2\eta}}\over
{ 16 \pi^2}} &&{{\eta^2\,\Gamma^4(\frac\eta2)}\over
{\Gamma(\eta)\,\Gamma(\eta+2)}} \nonumber \\
\times T^{-1+2\eta}\,\, &&\Theta(\eta, \omega/4\pi
T),
\end{eqnarray}
where the function $\Theta$ is
\begin{equation}
\Theta(\eta,x)={{\sinh(2\pi x)}\over{2\pi x}}
{{|\Gamma(\frac\eta2+i x)|^4}\over{\Gamma^4(\frac\eta2)}}\;
{{\eta^2+4 x^2}\over{\eta^2}}. \label{tfun}
\end{equation}
This  function is plotted on Figure~\ref{fig:hallcon} for
different values of $\eta$.

\begin{figure}
\centerline{\epsfig{file=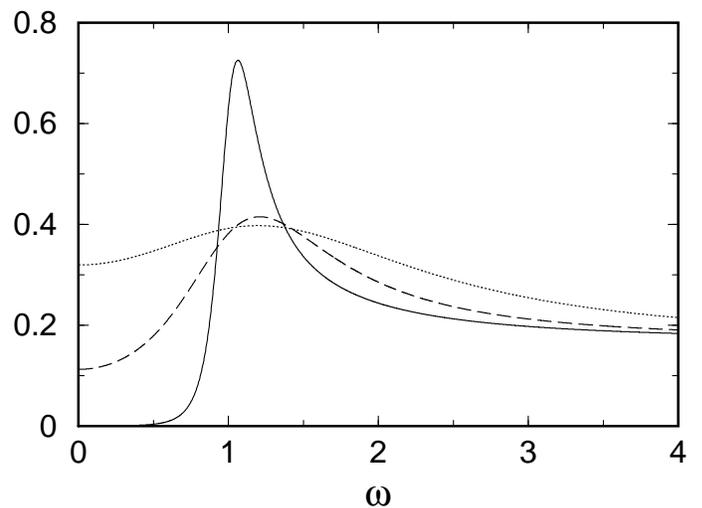,angle=0,width=\figwidth}}
\caption{\label{fig:hallrecon} The function $S(\eta,\omega,T)$
that determines the real part of the conductivity is plotted  for
$\eta=0.5$ and $h=1$. The solid line corresponds to $T=0.05$,
the dashed line to $T=0.2$ and the  dotted line to $T=0.4$.}
\end{figure}

In the presence of the magnetic field we shall plot the
dependence of the real part of the conductivity on frequency only
for a particular case $\eta=0.5$ (The value of $\eta$
realized in quasi-one-dimensional organic conductors is believed
to be close to $0.5$.) Presenting the conductivity as
\begin{equation}
 \rsig0 = {{a_y e^2 t_\perp^2(2\pi a_x)^
{2\eta}}\over{ 2\pi^2}} S(\omega)
\end{equation}
we plot the function $S$ on Figure~\ref{fig:hallrecon} for
different values of temperature. At low enough temperatures we
see a definite resonance at $\omega=h$. Also at low
temperatures we see the  effect that was mentioned above (see end
of Section~\ref{zerotem}): At frequencies less than $h$ the
dissipation is suppressed by the magnetic field. At higher
temperatures the resonance becomes smeared and eventually
disappears.

\section{Summary and Conclusion}
\label{conclusion}

The main result of the present paper is that, in the absence of
in-chain momentum relaxation, the Hall resistivity of weakly
coupled Luttinger chains is not affected by the interactions.
$\rho_{xy}$ is independent of frequency or temperature, and
simply given by the free fermion expression (\ref{rhoyx}) (see also
(\ref{cosine})). This result was established at zero-frequency
using Galilean invariance in the chains, and generalized at
finite frequency using exact relations based on current
commutation relations. This method also allows us to fully
determine the conductivity tensor in terms of a single quantity:
the inter-chain frequency dependent resistivity
Eqs.~((\ref{conduc_tensor})).

We have explicitly evaluated this quantity to lowest order in the
interchain hopping, as a function of frequency, temperature and
magnetic field. This expansion is valid in the Luttinger liquid
regime, i.e at high enough temperature $T\gg\epsilon^*$ or
frequencies $\omega\gg\epsilon*$, or at large magnetic field.
Various limiting forms of the general expression
(\ref{general_trans}) have been investigated, which reflect the
non-Fermi liquid aspects of the in-chain physics. Explicit
expressions of the associated scaling functions of $\omega/T$ and
$H/T$ have been given.

Because our calculation does not take into account the in-chain
momentum relaxation processes, it is rather difficult to make
definite statements concerning comparisons to the recent Hall
measurements on the quasi one-dimensional organic conductor \SP
\cite{moser_hall_1d,mihaly_hall_1d}. We note that both measurements 
yield values of $R_H$
close to the non-interacting band value at high enough
temperature, in qualitative agreement with our result. As
demonstrated in the present paper, this does not preclude strong
electron-electron interactions to be present in the chains.
Furthermore, it was shown in Ref.~\onlinecite{georges_organics_dinfiplusone},
using a generalized dynamical mean-field treatment, that the model of coupled
Luttinger liquid chains considered here can lead, in the low
temperature regime $T\ll\epsilon^*$ where Fermi liquid coherence
has set in, to a very small Drude weight. Our results therefore
show that there is no contradiction between this experimental
observation and the fact that $R_H$ is close to the band value.

The two recent experimental studies of the Hall effect in \SP
differ by the temperature dependence observed at high
temperature. In Ref.~\onlinecite{mihaly_hall_1d}, with the magnetic field 
parallel to
the chains, very little temperature dependence was observed. In
Ref.~\onlinecite{moser_hall_1d}, with the field along the least conducting 
axis, a
significant temperature dependence was measured, with $R_H(T)$
increasing by almost a factor of two between $150 K$ and $300 K$
(where a saturation is apparently reached). Obviously, the
temperature dependence of $R_H$ can only be addressed theoretically
once in-chain momentum relaxation processes are included. We have
not yet performed a detailed calculation along these lines, but
we conclude this paper by making a few general remarks about what
can be expected.

Momentum relaxation can be assigned to the inclusion of an
additional operator in the Hamiltonian of the chains, in addition to
the ordinary LL Hamiltonian. Let us consider the regime in which
this operator can be treated in perturbation theory.
Specifically, we have in mind for example the case of a
commensurate filling, where an umklapp operator is generated.
Perturbation theory is valid either when the umklapp is
irrelevant or, if it is relevant, when temperature is high enough
compared to the energy scale $\Delta_g$ associated with the
perturbation (i.e the induced Mott gap). For a model with spin at
commensurability $1/2n$ ($n=1$ for half-filling, $=2$ for
quarter-filling), we have $\Delta_g\sim g^x_3 $ with
$2x=1/(1-n^2K_{\rho})$. $g_3$ is the coupling constant and the
umklapp is relevant for $x>0$. For $g_3=0$, each chain is a perfect
conductor. For a non-vanishing but arbitrary small $g$, the
in-chain conductivity (in zero field) obeys the scaling behavior:
\begin{equation}
\sigma(\omega,T) = {1\over\omega}\,
\widetilde{\sigma}\left(\frac{g_3}{\omega^{1/x}},
\frac{g_3}{T^{1/x}}\right)
\end{equation}
with $\widetilde{\sigma}$ a universal scaling function (taking in
general complex values). This expression can be justified from
very general scaling arguments, observing that the conductivity
is related to the current correlation function by $\sigma\sim
\langle j.j \rangle/\omega$ and has thus the dimension of an inverse energy
($v_F=1$),
while $\omega/\Delta_g\sim \omega/g_3^x$ and $T/\Delta_g\sim T/g_3^x$
are the dimensionless scaling variables. It has also been
established by a memory function calculation
\cite{giamarchi_umklapp_1d,giamarchi_attract_1d}, which
allows an approximate determination of the scaling function
$\widetilde{\sigma}$. Let us recall two important limiting
behaviors of this expression. At high-frequency and low
temperature ($\omega\gg\Delta_g\gg T$), the real part
$\omega\sigma^\prime$ becomes a scaling function of
$g_3/\omega^{1/x}$ only, which vanishes for small arguments (since
at $g_3=0$ one has a perfect conductor and hence $\sigma$ is
imaginary). This scaling function has a regular Taylor expansion
in powers of $g/\omega^{1/x}$ (which starts at second order).
Hence the above scaling expression allows to simply predict the
dominant high-frequency behavior to be:
\begin{equation}
\sigma^\prime(\omega\gg\Delta_g) \sim
\frac{1}{\omega}\left(\frac{g_3^2}{\omega^{2/x}} + \cdots \right)
\end{equation}
Noting that $1+2/x=5-4n^2 K_{\rho}$, this is the result derived in
Ref.~\onlinecite{giamarchi_umklapp_1d}. In the opposite limit of low 
frequency, $T\sigma_{dc}$
becomes a scaling function of $g/T^{1/x}$, which obviously must
diverge at small $g_3$. Not surprisingly, it is the scaling
function associated with its inverse, the resistivity, which has
a smooth Taylor expansion in this limit, yielding:
\begin{equation}
\rho_{dc} = T \, \widetilde{\rho}\left(\frac{g_3}{T^{1/x}}\right)
\sim T \left(\frac{g_3^2}{T^{2/x}}+\cdots \right)
\end{equation}
Again, with $1-2/x=4n^2K_\rho-3$, this agrees with the result of
Ref.~\onlinecite{giamarchi_umklapp_1d}.

It is natural to attempt a generalization of these scaling
arguments to the Hall response. We focus here on the low-field
Hall number $R_H$ obtained from the linear term in the dc-Hall
resistance $\rho_{xy}=H\,R_H(T)+\cdots$. 
The dimensional arguments lead to the following expression 
for the Hall number
\begin{equation}
R_H =R^{(0)}_H \widetilde{R}\left(\frac{g_3}{T^{1/x}}\right),
\end{equation}
 where $R_H^0$ is a band value of the Hall resistivity
and $\tilde R$ is a dimensionless scaling function.
For small $g_3$ (or large $T\gg\Delta_g\sim g_3^x$), it is expected
from the results of the present paper that $R_H$ tends to the
band value $R_H^0$ (Galilean invariance is restored in this
limit). Since $R_H$ is obtained from a resistivity, it is natural
to expect that the scaling function $\widetilde{R}$ has a smooth
Taylor expansion in powers of $g_3/T^{1/x}$, and therefore that the
first corrections describing the deviation from saturation as
temperature is lowered are given by:
\begin{equation}
R_H(T) \,=\, R_H^0 \left(1+a_1 \frac{g_3}{T^{1/x}} + a_2
\frac{g_3^2}{T^{2/x}} +\cdots \right)
\end{equation}
It would be very interesting to confirm this expectation from an
explicit calculation and to determine the first coefficients in
this expansion. Note also that these considerations suggest that
a plot of $R_H(T)$ versus $\rho_{xx}/T$ should define a universal
scaling curve independent of the compound or external parameters
such as pressure (as long as $T\gg\Delta_g,\epsilon^*$). We hope
to address these issues in a forthcoming work, together with a
comparison of these scaling ideas to experiments.

\begin{acknowledgments}
A.G and T.G would like to thank Nancy Sandler, for discussions and
collaboration at an initial stage of this project, as well as D.
Jerome and J. Moser for thorough discussions of their
experimental results prior to publication. A.G also acknowledges
fruitful discussions with J.Cooper, L. Forro, A. Sengupta and V. Yakovenko.
A.L. would like to thank  L. Ioffe,  A.J. Millis and V. Yakovenko for
useful discussions. 
The authors are grateful
to the Institute for Theoretical Physics (Santa Barbara) which
partially supported this research (under NSF grant No. PHY94-07194).
\end{acknowledgments}

\appendix

\section{Explicit calculation of the high-frequency Hall
conductivity} \label{hallcon}

Considering the Hall conductivity we shall assume that the
electric field is applied along the chains (x-axis) $E_x=E_0
e^{-i\omega t},$ thus the Hall conductivity $\sigma_{xy}$ relates
the transverse current $j_y$ with electric field $E_x$
\begin{equation}
j_y(\omega)=\sigma_{yx}(\omega)\,E_x(\omega).
\end{equation}
Using Kubo formula and the perturbation theory in the hopping term
$t_\perp$ for the Hall conductivity we get\cite{lopatin_hall_luttinger}
\begin{equation}
\sigma_{yx}(\omega)=-{{2eit_\perp^2 }\over \omega} \left[\int
_0^\beta d\tau \,e^{i\Omega \tau}\, \Gamma(\tau)
\right]_{\Omega\to -i\omega} \label{sigma},
\end{equation}
where $\Gamma(\tau)$ is
\begin{eqnarray}
\Gamma(\tau_1-\tau_3) =\sum_{s=\pm}\int dx_3 dx_2 d\tau_2
\Bigl[\langle \psi_s(x_2,\tau_2)\psi_s^\dagger(x_1,\tau_1)\rangle \nonumber \\
\langle j(x_3,\tau_3)\psi_s(x_1,\tau_1)
\psi_s^\dagger(x_2,\tau_2)\rangle
+(x_1,\tau_1\leftrightarrow x_2,\tau_2)\Bigr] \nonumber
\end{eqnarray}
\begin{equation}
\times\sin h(x_1-x_2) \label{gamma}.
\end{equation}
The angular brackets in (\ref{gamma}) represent averaging
with respect to the single chain Lagrangian with density
\begin{eqnarray}
{\cal L}^{(1)}= \psi^\dagger(-\partial_\tau+iv_F\,\tau_3
\,\partial_x&+&\alpha\,
\partial_x^2)\psi  \nonumber \\
&-&g\,\psi_+^\dagger\psi_+\psi^\dagger_-\psi_-, \label{singlelag}
\end{eqnarray}
and the current $j$ is the single-chain  current
\begin{equation} \label{eq:singlecurrent}
j=ev_F\;\psi^\dagger\tau_3\psi +2 e \alpha\,
\psi^\dagger(-i\partial_x)\psi. \label{current}
\end{equation}
Note that it contains an $\alpha-$contribution arising from the
nonlinear correction to the spectrum.

The single-chain correlation functions in (\ref{gamma}) can be calculated
by the bosonization method. The functional technique allowing to
find the necessary correlation functions at zero temperature was
described in the previous work of one of 
authors\cite{lopatin_hall_luttinger}.  The generalization to finite
temperatures is straightforward and in the following we will
summaries only the results.
 The single-chain Lagrangian (\ref{singlelag})
in the bosonized form is
\begin{equation}
{\cal L}^{(1)}={\cal L}^{(0)}+V(\Pi,\partial_x\Phi) \label{lone},
\end{equation}
where
\begin{eqnarray}
{\cal L}^{(0)}=
i\,\Pi\,\partial_\tau\Phi-{1\over 2}\left(\Pi^2+(\partial_x\Phi)^2\right),\\
V(\Pi,\partial_x\Phi)=-{\alpha\over
3}\tilde\beta\,\partial_x\Phi\left(3{\pi\over{\tilde\beta^2}}\Pi^2+
{{\tilde\beta^2}\over\pi}(\partial_x\phi)^2\right),
\end{eqnarray}
where $\tilde\beta$ is related with $\eta$ by $2\eta=
\tilde\beta^2/\pi+ \pi/\tilde\beta^2-2.$

The bosonic fields $\Phi,\Pi$ and original ``fermionic'' $\psi,\psi^*$ fields
are related by
\begin{eqnarray}
\psi_{\pm}(x,\tau)={1\over{\sqrt{2\pi a_x}}}e^{\pm i\Phi_\pm(x,\tau)}, \\
\Phi_\pm(x,\tau)=\tilde\beta\Phi(x,\tau)
\mp{\pi\over\tilde\beta}\int_{-\infty}^x
dx^\prime\Pi(x^\prime,\tau).
\end{eqnarray}
 As was pointed out in Ref.~\onlinecite{lopatin_hall_luttinger}, the Green function
calculated by the functional method
\begin{equation}
G_f(x_1-x_2,\tau_1-\tau_2)=\langle
\psi(x_1,\tau_1)\psi^*(x_2,\tau_2)\rangle
\end{equation}
must be multiplied by ${\rm sgn}(\tau_1-\tau_2)$ to restore the
proper symmetry of the Green function arising from the fermionic
anticommutation relations.  The Green function corresponding to
the Lagrangian ${\cal L}^{(0)}$ was already  written above
(\ref{tgreenf}).
Since the interaction $V$ will be treated as a perturbation it is very
convenient to introduce the following generating functional
\begin{equation}
Z_\pm(f_0,f_1)=\langle \psi_\pm(\xi_1)\psi_\pm^*(\xi_2) e^{ \int
d^2\xi\left[f_0(\xi)\Pi(\xi)+f_1(\xi)\partial_x\Phi(\xi)\right]}
\rangle^{(0)}, \label{z0}
\end{equation}
where $\xi=(x,\tau)$ and $d^2 \xi=dx\, d\tau.$
The straightforward calculation gives
$$
\hspace{-2.6cm}
Z_{\pm}(f_0,f_1)=G^{(0)}_{\pm,f}(\xi_1,\xi_2)
$$
\vspace{-0.6cm}
\begin{equation}
\times e^{F^{\pm}(f_0,f_1)+{1\over 2} \int d^2 \xi_1d^2\xi_2
f^T(\xi_1)D(\xi_1,\xi_2)f(\xi_2)}, \label{zf}
\end{equation}
where  $G^{(0)}_{\pm,f}$ is the Green function (\ref{tgreenf})
multiplied by ${\rm sgn}(\tau_1-\tau_2)$, $f=(f_0,f_1)$ is a
two-vector constructed from $f_0,f_1$, and
 $F^{\pm}(f_0,f_1)$ is a linear functional of $f_0, f_1$
\begin{eqnarray}
F^{\pm}(f_0,f_1)= \int d^2\xi
&&\Bigl(f_0(\xi)J_0^{\pm}(\xi_1,\xi_2,\xi) \nonumber \\
&&+f_1(\xi)J_1^{\pm}(\xi_1,\xi_2,\xi) \Bigr) ,
\end{eqnarray}
with
\begin{eqnarray}
&&J_0^{\pm}(\xi_1,\xi_2,\xi_3)= i {T\over 4}\Biggl[
\left(\mp\tilde\beta-{\pi\over{\tilde\beta}}\right)\coth(z_1-z_3)\pi T
\nonumber \\
&&+\left(\pm \tilde\beta-{\pi\over{\tilde\beta}}\right)
\coth(z_1^*-z_3^*)\pi T
-(\xi_1\leftrightarrow\xi_2)      \Biggr], \label{j1} 
\end{eqnarray}
\begin{eqnarray}
J_1^{\pm}&&(\xi_1,\xi_2,\xi_3)=i {T\over 4}\Biggl[
\left(\pm\tilde\beta+{\pi\over{\tilde\beta}}\right)\coth(z_1-z_3)\pi T  \nonumber \\
&&+\left(\pm \tilde\beta-{\pi\over{\tilde\beta}}\right)
\coth(z_1^*-z_3^*)\pi T -(\xi_1\leftrightarrow \xi_2) \Biggr]
\end{eqnarray}
where $z=x+i\tau,z^*=x-i\tau.$
 Finally, $D$ is a matrix  Green function which in the momentum
space is
\begin{equation}
D(\omega,p)={{p^2}\over{p^2+\omega^2}}\left[ \begin{array}{cc}
1&i{\omega\over p}\\i{\omega\over p}&1
\end{array} \right]. \label{gd}
\end{equation}

Up to the first order in the nonlinear correction $\alpha,$ the
expression for $\Gamma$ (\ref{gamma}) can be schematically
presented as
\begin{equation}
\langle j\psi\psi^\dagger\rangle\langle\psi\psi^\dagger\rangle=
\langle j_0\psi\psi^\dagger\rangle\langle\psi\psi^\dagger\rangle+
\langle j_1\psi\psi^\dagger\rangle^{(0)}\langle
\psi\psi^\dagger\rangle^{(0)},\label{pre}
\end{equation}
where $j_0$ and $j_1$ correspond to the first and second terms in
the equation for the current (\ref{current}) respectively. The
first term in (\ref{pre}) contains the current to the
 zeroth order in $\alpha$ so that the nonlinear corrections
come from the Lagrangian. The second term contains no
$\alpha$-corrections from the Lagrangian because $j_1$ is already
proportional to $\alpha$. It was shown in
Ref.~\onlinecite{lopatin_hall_luttinger} that the first term
gives no contribution to the Hall conductivity. One can check that
the same holds for finite temperatures. Thus we need to find only
the contribution from the second term. The correlator $$ \int
dx_3 \langle
j_1(\xi_3)\psi_{\pm}(\xi_1)\psi_{\pm}(\xi_2)\rangle^{(0)}, $$
where the current correction $j_1$ in the bosonized form is $$
j_1=-2e\alpha\,\Pi\,\partial_x\Phi, $$
 can be calculated with the help of the generating functional using
Eqs.(\ref{z0},\ref{zf}) $$ \int dx_3\langle
j_1(\xi_3)\psi_{\pm}(\xi_1)\psi_{\pm}(\xi_2)\rangle^{(0)} $$ $$
=-2e\alpha\int dx_3
J_0^\pm(\xi_1,\xi_2,\xi_3)J_1^\pm(\xi_1,\xi_2,\xi_3)
G_{\pm}^{(0)}(\xi_1,\xi_2), $$ where we used that $\int dx
D(x,\tau)=0.$ To obtain the Hall conductivity (\ref{sigma}) we
first need to take the integrals over  $x_3$ and $\tau_3$ in
\begin{eqnarray}
\int dx_3\,&& d\tau_3\, e^{i \Omega \tau_3 } \,\langle
j_1(\xi_3)\psi_\pm(\xi_1)\psi^\dagger_\pm(\xi_2)\rangle
\nonumber \\
&&={{e\alpha T}\over{2\Omega}}G_\pm(x_2,t_2)
\Biggl[\left(\pm\beta-{\pi\over\beta}\right)^2\coth(z_2^*\pi T) \nonumber \\
+&&\left(\pm\beta+{\pi\over\beta}\right)^2 \coth(z_2\pi T)
\Biggr] \,(e^{-i\Omega \tau_2}-1). \label{corj}
\end{eqnarray}

Using Eq.~(\ref{corj}) and the relation
$2\eta=\tilde\beta^2/\pi+\pi/\tilde\beta^2-2$ for the function
$\Gamma$ we get
\begin{eqnarray}
\Gamma &&(\Omega)={{8\pi e \alpha T}\over{\Omega }}\int_0^\beta
d\tau\,dx\, G_+^2(x,\tau) \Bigl[\frac\eta2\coth[(x-i\tau)\pi T]
\nonumber  \\  
&&+(\frac\eta2+1) \coth[(x+i\tau)\pi T] \Bigr]
\sin(h\,x) \Bigl[e^{i\tau\Omega}-1\Bigr] \label{gammafin}
\end{eqnarray}

Using the explicit expression for the Green function
(\ref{tgreenf}) and integrating (\ref{gammafin}) over $x$ by
parts we get
\begin{eqnarray}
\sigma_{yx}(\omega)={{8 t_\perp^2 e^2 h\alpha }\over{\omega^2}}
\Biggl[
\int dx \int_0^{\beta} d\tau\; e^{i\Omega \tau}\; G_+^2(x,\tau) \nonumber \\
\times \cos(h x) -(\Omega=0)\Biggr]_{\Omega\to-i\omega}. \label{hfhallcond}
\end{eqnarray}
 To the lowest order in the curvature $\alpha$ and the interchain hopping
$t_\perp$ the conductivity along the chains is given by
\begin{equation}
\sigma_{xx}^{(0)}=-{{v_F\,e^2}\over{a_y\,\pi}} {1\over{i\omega}}.
\label{sxx}
\end{equation}
In addition, keeping only the leading order in the hopping $t_\perp$
we have
\begin{equation}
\rho_{xy}=- {{\sigma_{xy}}\over{\sigma_{yy}\sigma_{xx }}}.
\label{rsigma}
\end{equation}
since $\sigma_{yx} \sim t_\perp^2$ at high enough frequency.
Using (\ref{rsigma},\ref{sxx},\ref{hfhallcond},\ref{trcond}) 
 we calculate the Hall resistivity obtaining
the simple result
\begin{equation}
\rho_{yx}=2\pi\alpha\, {{H a_y}\over{v_F e c}} \label{result}.
\end{equation}

\section{Expansion of the inter-chain resistivity in the band curvature}
\label{ap:expan}

In this appendix, we show that the inter-chain resistivity has a
well-behaved expansion in powers of the band curvature parameter
$\alpha$.
We found that at low $\omega$ the conductivity $\sigma_{yy}$
behaves as
\begin{equation}
\sigma_{yy}(\omega)=
-{{v_F e^2}\over{\pi a_y}} {{i\omega}\over{\gamma^2}}, \;\;\;\;
\omega\to 0.
\end{equation}

 This term becomes of the order of $\sigma_{yy}^{(0)}(\omega=0)$
at the frequency of order of $\omega_0$ defined by (\ref{omegazero}).
Therefore the next term in the expansion should have form
\begin{equation}
\sigma_{yy}(\omega)=
-{{v_F e^2}\over{\pi a_y}} {{i \omega}\over{\gamma^2}}\left[
1+ \kappa {{\omega i}\over{\omega_0}} \right] , \;\;\;\;\;\;\; 
\omega\to 0 \label{expan}
\end{equation}
where $\kappa$ is a coefficient of order 1.
Now using the expansion (\ref{expan}) and formula (\ref{conduc_tensor})
for the resistivity at $\omega\to 0$ we get
\begin{equation}
\rho_{yy}(\omega)=\sigma_{yy}^{-1}(\omega)+
{{\gamma^2}\over{i\omega}} {{a_y \pi}\over{v_F e^2}} \to \kappa/
\sigma_{yy}^{(0)}(\omega=0).
\end{equation}
 Thus the resistivity $\rho_{yy}(\omega)$ is regular in $\alpha$ 
even at low frequency. Also we expect that $\kappa=1$
because the effect of curvature must be always small 
(as a ratio of the effective energy of the problem and Fermi energy)
for quantities that have a regular dependence on $\alpha,$
so the zeroth order (in $\alpha$) answer should give the correct result
for the transverse resistivity $\rho_{yy}.$

\section{Calculation of the real part of the transverse conductivity}
\label{appx_trans}

\begin{figure}[here]
\unitlength1.0cm
\begin{center}
\begin{picture}(7.5,6)
\epsfysize=6.0 cm \epsfxsize=7.5 cm \epsfbox{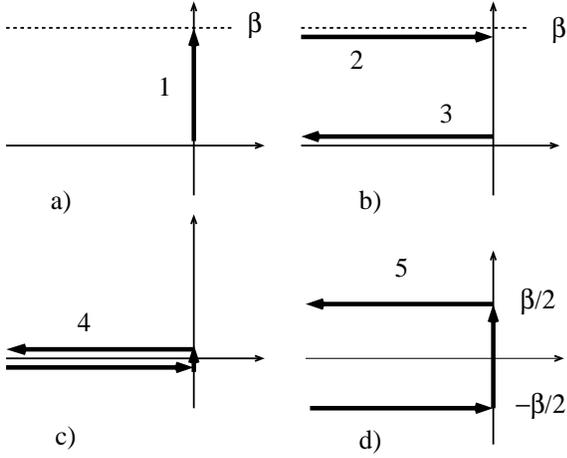}
\end{picture}
\vspace{0.5cm} \caption{This figure shows the transformations of
the contour integrals allowing to find the real part of the
conductivity (\ref{trcond}).}
\end{center}
\label{contour}
\end{figure}

The real part of the conductivity (\ref{trcond}) can be
calculated analytically. To do this, it is convenient to
introduce a complex variable $t=i\tau$ in Eq.(\ref{trcond}), so
that the integral over $\tau$ is represented by the contour 1 on
Fig.5a. The Green function (\ref{tgreenf}) as a function of
$t=i\tau$ has poles and branch cuts only along the lines
$t=i\beta n,$ where  $n$ is an integer. Therefore the integral
along the contour 1 is equal to the integrals over contours 2 and
3 (Fig. 5b). Shifting the contour 2 on $-i\beta $ (this can be
done because $G^2(x,t+i\beta)=G^2(x,t)$ and $\Omega$ is an even
Matsubara frequency) we represent the conductivity as an integral
over contour 4 (Fig 5c). At this step we can make an analytical
continuation $\Omega\to-i\omega,$ because the integral over $t$
runs  over real negative values. After the analytical
continuation is done, it is convenient do deform the contour 4
into contour 5 (Fig 5d). One can see that the integrals over
parts of the counter 5 that are parallel to the real axis
determine the real part of conductivity
$$
\sigma^{\prime(0)}_{yy}(\omega)=-{2{a_y\,e^2t_\perp^2T^2(\pi T
a_x)^{2\eta}} \over{\omega}}\int_{-\infty}^0 dt\int dx\,
e^{-i\omega t} 
$$ \vspace{-0.3cm}
\begin{equation} \times {{[\cosh[(x+t)\pi T]\cosh[(x-t)\pi T]]^{-\eta }}\over
{\cosh^2[(x+t)\pi T]}} \sinh{\omega\over {2T}}. \label{re}
\end{equation}
 The part of the contour integral 5 which is along the imaginary axis gives
the imaginary part
\begin{eqnarray}
\sigma_{yy}^{\prime\prime(0)}(\omega)={{a_y\,e^2t_\perp^2T^2(\pi T
a_x)^{2\eta}}
\over{\omega}} \int_{-\beta/2}^{\beta/2} dt\int dx \nonumber \\
{{[\sinh[(x+it)\pi T] \sinh[(x-it)\pi T]]^{-\eta}} \over
{\sinh^2[(x+it)\pi T]}} (e^{\omega t}-1). \label{im}
\end{eqnarray}
 The integral over t in Eq.(\ref{re}) can be extended to run from
$-\infty$ to $\infty$ because the expression under the integral
is even in $t.$ After this one can  take the integrals over $x$
and $t$ introducing  the new variables $x-t$ and $x+t$ and get
the answer
\begin{eqnarray}
\sigma^{\prime(0)}_{yy}(\omega)&=&{{ a_y e^2 t_\perp^2(2\pi T
a_x)^{2\eta}}\over {2\pi^2}} {{\sinh (\omega/2T)}\over
{\omega}}  \nonumber {{(\frac\eta2)^2+{{\omega^2+h^2}\over{(4\pi
T)^2}} }\over {\Gamma(\eta)\Gamma(2+\eta)}}
\nonumber \\
&&\left|\Gamma\left(\frac\eta2+{{\omega+h}\over{4\pi T}}i\right)
\right|^2 \left|\Gamma\left(\frac\eta2+{{\omega-h}\over{4\pi
T}}i\right) \right|^2.
\end{eqnarray}


\end{document}